\documentclass{aa}

\usepackage[latin1]{inputenc}     
\usepackage[english]{babel}       
\usepackage{amssymb}              
\usepackage{amsmath}
\usepackage{amsfonts}             
\usepackage{graphicx}             
\usepackage{epstopdf}             

\usepackage{natbib}
\usepackage{epsfig}
\usepackage{txfonts}

\begin{document}

\title{Extracting cosmological signals from foregrounds in deep mm maps of the sky}

\author{Luca Conversi\inst{1,2}, Paola Fiadino\inst{2}, Paolo de Bernardis\inst{2,3}, Silvia Masi\inst{2,3}}

\institute{European Space Astronomy Centre, European Space Agency, P.O. Box 78, 28691 Villanueva de la Ca\~{n}ada (Madrid), Spain
	\and Dipartimento di Fisica, Universit\`{a} di Roma ``La Sapienza", P.le A. Moro 2, 00185 Roma, Italy
	\and INFN Sezione di Roma c/o Dipartimento di Fisica, Universit\`{a} di Roma ``La Sapienza"
}

\offprints{luca.conversi@esa.int}
\date{Submitted: xxx; Accepted: xxx}

{\abstract
	{The high Galactic latitude sky at millimeter and submm wavelengths contains significant cosmological information about the early Universe (in terms of the cosmic microwave background) but also the process of structure formation in the Universe from the far infrared background produced by early galaxies and the Sunyaev-Zeldovich effect in clusters of galaxies.}
	{As the Planck mission will produce full sky maps in this frequency range, deeper maps of selected low-foregrounds patches of the sky can produce complementary and important information. Here we analyze the performance of a balloon-borne survey covering a $10^\circ \times 10^\circ$ patch of the sky with a few arcminute resolution and very high pixel sensitivity.}
	{We simulate the different components of the mm/submm sky (i.e., CMB anisotropies, SZ effect, radio and infrared sources, far infrared background, and interstellar dust) using current knowledge about each of them. We then combine them, adding detector noise, to produce detailed simulated observations in four observational bands ranging from 130 to 500 GHz. Finally, we analyze the simulated maps and estimate the performance of the instrument in extracting the relevant information about each of the components.}
	{We find that the CMB angular power spectrum is accurately recovered up to $\ell \sim 3000$. Using the Sunyaev-Zel'dovich effect, most of the galaxy clusters present in our input map are detected (60\% efficiency overall). Our results also show that much stronger constrains can be placed on far infrared background models.}
	{}
}


\authorrunning{Conversi \emph{et al.}}
\titlerunning{Extracting cosmological signals from deep mm maps of the sky}
\maketitle

\section{Introduction}\label{introduction}

The high Galactic latitude sky at far infrared and millimetric frequencies is mainly dominated by four components: the cosmic microwave background (CMB), the Sunyaev-Zel'dovich (SZ) effect in clusters of galaxies, Galactic dust, and the far infrared background (FIRB) from unresolved point sources, as well as radio and infrared sources. The ``cosmological window'' extends roughly from 70 to 200~GHz, consisting at lower frequencies of the interstellar emission of spinning dust grains, free-free and synchrotron emission from the interstellar medium, which dominate over the cosmological microwave background, and at higher frequencies of the clumpy foreground from ``cirrus clouds'' of interstellar dust and the background signal due to unresolved sources at high redshift, which dominate over the sky brightness.

The CMB provides, together with its anisotropies, significant information on the primordial phases and the evolution of the Universe. Its frequency spectrum can be distorted by different effects \citep{2008RPPh...71f6902A}, occurring on both local or global scales: in this work, we only consider the local distortion due to the Sunyaev-Zel'dovich effect caused by the interaction between the hot gas present in the clusters of galaxies and the photons of the cosmic background radiation. To allow the extraction of the cosmological parameters, we need to separate these different types of emission, hence to use multi-band experiments \citep{2007A&A...465...57J}. Moreover, the study of these foreground signals is very useful to improve our knowledge about structure formation of the primordial Universe \citep{2000A&A...360....1G, 2005A&A...439..901A, 2006A&A...455..741P}.

In this paper, we develop an accurate simulation of the expected signals in the $130 - 500$~GHz frequency range, and demonstrate how it is possible to separate and recover the input signals. We use data from a multi-band, high resolution survey of $10^\circ \times 10^\circ$ of the sky, which is feasible using a long-duration balloon telescope. As baseline, we assume the characteristics of the OLIMPO experiment. The paper is organized as follows: in Sect.~\ref{experiment} we briefly summarize the characteristics of OLIMPO; in Sect.~\ref{model} we describe the astrophysical components included in our model; Sect.~\ref{map_generation} explains how we modeled each component and the map-making process; Sect.~\ref{map_extraction} describes the map-separation algorithm we used; finally, we discuss our results in Sect.~\ref{results} and report our conclusions in Sect.~\ref{conclusions}.

\section{The OLIMPO experiment}\label{experiment}

OLIMPO is a balloon borne experiment \citep{2005erbp.conf..581M} whose purpose is to observe selected portions of the sky on four frequency bands centered on $\nu = 143$, 217, 353, and 450~GHz, with an angular resolution ranging from $5'$ FWHM at 143~GHz to $2'$ FWHM at 450~GHz (see Table~\ref{tab:olimpo}). It is constituted by a Cassegrain telescope of 2.6~m of diameter, advanced bolometric detectors, and a long duration $^{3}$He cryostat, which will be mounted on a stratospheric balloon payload able to perform long duration ($10 - 15$ days) circumpolar flights.

The incoming radiation is focused onto four focal planes using a system of three dichroics: each focal plane consists of an array of TES (transition edge superconductor) detectors. The four arrays are composed of 19 detectors in the two lower frequency bands (143 and 217~GHz) and 24 in the others. Each array fills the optically correct area of the focal plane ($\sim 15'$ in diameter projected onto the sky). The cryogenic reimaging optics is mounted inside the experiment's cryostat, cooled to 2~K, while the bolometers are cooled to 300~mK: extensive baffling and a cold Lyot-stop reduce both straylight and sidelobes by about 30~dB. The optical system was optimized to allow diffraction-limited operations even with significant tilt of the primary mirror during sky scans. Radiation is refocused into the cryogenic optics through a side of the cryostat, to allow operation of the system with the telescope axis horizontal or at low elevation, during ground-based calibrations or in-flight observations of planets.

The high angular resolution and the wide range of frequencies covered by OLIMPO allow us to achieve the following scientific goals:
\begin{itemize}
	\item measurement of the CMB anisotropies at high multipoles ($\ell \gtrsim 2000$): by taking advantage of its wide frequency coverage and the high sensitivity, OLIMPO will significantly improve our current knowledge, allowing the study of the damping tail of the angular power spectrum of the CMB;
	\item measurement of the Sunyaev-Zel'dovich effect caused by cluster of galaxies: it is not possible to study the characteristics of high redshift clusters using X-ray data, since the signal is too faint to be detected in a blind survey with the present X-ray satellites. The Sunyaev-Zel'dovich effect, instead, is independent of the redshift of the cluster, thus a fundamental tool for cosmological measurements. The survey of known clusters and the deep blind observation of a blank sky area with OLIMPO will permit a more complete statistical analysis of the clusters distribution and of their evolution;
	\item study of the far infrared background (FIRB) due to unresolved galaxies at high redshift: this measurement will improve our knowledge of the galaxy formation process, and in general of the epoch immediately following recombination.
\end{itemize}

Because of its larger telescope, OLIMPO has a higher angular resolution than Planck HFI in the same bands \citep{2003NewAR..47.1017L}. Moreover, working as an observatory, it can integrate longer than Planck on selected sky regions. This, combined to the larger number of detectors per band, results in a higher sensitivity for the selected targets.

OLIMPO also complements BLAST \citep{2008ApJ...681..400P} and Herschel SPIRE \citep{2010A&A...518L...3G}, covering lower frequency bands and thus allowing clearer component separation, especially for the SZ effect; on the other hand, it complements ACT \citep{2004SPIE.5498....1F} and SPT \citep{2004SPIE.5498...11R} at higher frequencies and with comparable angular resolution.

\begin{table}[htd]
\centering
\begin{tabular}{l|c|c|c|c}
	Channel								& 143	& 217	& 353	& 450 \\
	\hline
	\hline
	Central frequency [GHz]				& 146	& 221	& 360	& 458 \\
	Bandwidth [GHz FWHM]					& 40		& 61		& 98		& 126 \\
	Field of view radius [arcmin]		& 20		& 20		& 20		& 20 \\
	Resolution [arcmin FWHM]				& 5.2	& 3.7	& 2.3	& 1.9 \\
	Number of detectors					& 19		& 19		& 24		& 24 \\
	Detector's NEP [$\mu K / \sqrt{Hz}]$	& 137.2	& 240.7	& 316.4	& 533.5 \\
\end{tabular}
\caption{Characteristics of the OLIMPO instrument in the 4 wavebands. Filters' properties, detector's NEP, and resolution come from laboratory measurement of flight samples. Each detector array fills the same unvignetted, circular field of view.}\label{tab:olimpo}
\end{table}

\section{Model parameters}\label{model}

As described in Sect.~\ref{introduction}, the sky brightness in the frequency range $130 - 500$~GHz is mainly dominated by four components: cosmic microwave background, Sunyaev-Zel'dovich effect, Galactic dust, and far infrared background. In addition to these signals, both instrumental noise and the contribution of radio and infrared point sources must be counted. Hence, we built a model to describe the expected signals in four frequency bands (nominally centered on $\nu = 143$, 217, 353, and 450~GHz) to check the capabilities of the OLIMPO experiment, and whether it can separate the different effects and to what extent.

\subsection{Cosmic microwave background}

The cosmic microwave background (CMB, \citealt{penzias}) is the brightest background of the millimeter sky. This radiation and its properties are the strongest confirmations of the standard cosmological model \citep{1948Natur.162..680G}. It has a thermal black-body spectrum at a temperature of 2.725~K \citep{1999ApJ...512..511M}, thus the spectrum peaks at 160~GHz, while the anisotropy spectrum peaks at 220~GHz. At present, the more robust measurement available of the angular distribution of the CMB is that derived from the WMAP 5-year data \citep{2009ApJS..180..306D}: using its estimation of the CMB angular power spectrum, it is possible to generate CMB simulated maps as Gaussian random realizations of the Fourier transform of the CMB spectrum.

\subsection{Sunyaev-Zel'dovich effect}

The SZ effect is the result of the interaction between the CMB photons and the hot gas of electrons ($T_e \gtrsim 5$ keV) in clusters of galaxies by inverse Compton scattering, i.e.\ the photons adsorb part of the electron's energy: this leads to a blueshift of the CMB frequency spectrum \citep{1969Ap&SS...4..301Z, 1999PhR...310...97B}. Two contributions can be identified: a \emph{thermal} one, directly related to $T_e$, and a \emph{kinetical} one, which is related to the cluster peculiar velocity $V_r$.

In the non-relativistic limit, also known as the \emph{Kompaneets approximation} \citep{kompaneets}, the frequency spectrum of the thermal SZ effect is described by
\begin{equation}\label{eq:kompaneets}
	\frac{\Delta T}{T_{\mathit{CMB}}}=\left[ x \cdot \frac{e^{x}+1}{e^{x}-1} - 4 \right]
	\cdot y = \coth \left( \frac{x}{2} \right) \cdot xy - 4 ,
\end {equation}
where $x=h\nu / k_{B} T_{\mathit{CMB}}$ and the \emph{comptonization parameter} $y$
\begin{equation}\label{eq:compton}
	y = \sigma_{T} \int \frac{k_{B}\, T_{e}}{m_{e}\, c^{2}}\, n_{e}\, dl ,
\end{equation}
where the integration is performed along the line of sight. The thermal SZ effect is described by the product of the analytic expression of the spectrum and the comptonization parameter. Thus, in this approximation, the spectrum does not depend on the cluster's parameters.

The comptonization parameter is obtained from hydrodynamical simulations \citep{2005MNRAS.361..233B, 2005Natur.435..629S} that follow the process of formation and evolution of galaxy clusters in a cosmological framework. These simulations include an advanced treatment of several processes, such as radiative cooling, star formation, thermal conduction, release of metals, and energy feedback from supernovae type Ia and II. The result of these simulations is a box containing a very high number of particles ($\sim 10^{10}$ in the largest one, the \emph{Millennium simulation}, \citealt{2005MNRAS.364.1105S}): by integrating along the line of sight, it is possible to obtain an accurate simulation of a map of the comptonization parameter. 

As already mentioned, when a cluster has a non-zero peculiar velocity in the CMB reference frame, there is a kinematic effect in addition to the purely thermal SZ one. In the reference frame of the scattering gas, CMB radiation appears anisotropic, and the effect of the inverse Compton scattering is to slightly re-isotropize the radiation.

Assuming that the thermal and kinematic effects are independent, it is possible to calculate the kinematic correction to the spectrum
\begin{equation}\label{delta:i:kin}
   \Delta I_K (x) = -\frac{2\, (k_{B}\, T_\mathit{CMB})^{4}}{h^{3}\, c^{2}}
      \cdot \frac{V_{r}}{c} \cdot \tau_{e} \cdot \frac{x^{4}\, e^{x}}{(e^{x} - 1)^{2}} ,
\end{equation}
where $\tau_{e} = \sigma_T \int n_e\, dl$ is the optical depth. In principle, a photometric measurement at $x \simeq 3.83$ ($\nu \simeq 220$ GHz) is able to detect a non-zero cluster's peculiar velocity along the line of sight: given this parameter, it is possible to subtract the kinematic SZ signal at the other frequencies and distinguish the thermal component. Therefore, this effect can be used to measure the large-scale motion of objects at high redshifts. However, the spectrum of the kinematic SZ is identical to the spectrum of primary CMB anisotropy, so it is almost impossible to distinguish the two contributions along the same line of sight. Therefore, in the following, we do not consider this effect.

\subsection{Dust}\label{sfd_model}

The Schlegel, Finkbeiner and Davis model $\# 8$ \citep{1998ApJ...500..525S} is an accurate description of the far infrared and microwave emission from the diffuse interstellar dust in the Galaxy. It is based on high resolution, far infrared IRAS \citep{1994STIN...9522539W} data and calibrated with the DIRBE \citep{1992ApJ...397..420B} data at $\nu = 3000$~GHz, which have a much higher calibration accuracy but poorer spatial resolution. This model then extends to FIRAS \citep{1994ApJ...420..457F} and DMR \citep{1990ApJ...360..685S} frequencies, $100 - 2100$~GHz and $31.5 - 90$~GHz respectively. The outputs of the model are full sky maps at submillimeter and millimeter wavelengths.

This is a multicomponent model, because it takes into account an interstellar medium of many types of molecules and dust grains, and fits all available data in the range $200 - 3000$~GHz. The model analyzes only the sky regions where the far infrared emission is expected to be dominated by diffuse interstellar medium. Hence, a spatial mask that excludes all the undesirable regions is created, leaving a map covering $\sim 71 \%$ of the sky observed by FIRAS.

This model was developed for the limit of quite big grains, $0.01 < a < 0.25\, {\rm \mu m}$, where $a$ is the grain radius: they reach a steady thermal equilibrium with the surrounding radiation field. Furthermore, since the grain's emission wavelength scales proportionally to its volume $a^{3}$ , larger grains dominate the submillimeter emission. Smaller grains, $a < 0.01\ {\rm \mu m}$, can be heated temporarily to very high temperatures and dominate the emission at $\lambda < 100\ {\rm \mu m}$ ($\nu > 3000$~GHz), but they neither contribute significantly to submillimeter emission, nor are important in the FIRAS frequency interval.

\subsection{Far infrared background}

The cosmic infrared background (CIB) was first detected in COBE maps by \cite{1996A&A...308L...5P} and later endorsed by \cite{1996ApJ...473..576F}, \cite{1998ApJ...508..123F}, and \cite{1998A&A...334..420B}. SCUBA \citep{2006ApJ...644..769D}, Spitzer \citep{2007ApJ...665L..89L}, and BLAST \citep{2009Natur.458..737D} also confirmed its detection. The CIB is a product of the cosmological structure formation process: matter aggregates into stars and galaxies, then the evolution of these systems is related to gravitational and nuclear processes, which lead to the emission of energy. Cosmic expansion and absorption of radiation emitted by dust transform this radiative energy into CIB emission \citep{2001ARA&A..39..249H, 2002PhR...369..111B}.

One of the main problems of far infrared observations is the difficulty in resolving sources (\emph{confusion noise}): this is due to the low spatial resolution of the instruments, caused by diffraction in small telescopes. Thus, the superimposition of signals coming from faint unresolved sources form the so-called far infrared background. Measurements obtained by BLAST \citep{2009Natur.458..737D} have been less affected by this problem because this telescope was able to resolve most of the sources of the CIB.

Different models have been developed to reproduce the FIRB measurements (e.g., \citealt{1994ApJ...427..140F, 1996MNRAS.283..174P, 1997MNRAS.287L..17B, 1997ESASP.401..159F, 1998MNRAS.295..877G, 2001ApJ...562..179X, 2009ApJ...701.1814V}). We adopted the model of Lagache, Dole \& Puget (hereafter LDP, \citealt{2003MNRAS.338..555L}), which has the advantage of being simple and with a low number of parameters and components, but is cabable of reproducing all the available observations. In a simple way, it constrains the galaxy luminosity function evolution with redshift, and fits the existing source number counts and redshift distributions, and the CIB intensity and its fluctuations, from the mid to the far infrared wavebands \citep{2008A&A...481..885F}. This model is based on both local spectral templates of starburst and normal galaxies and the local infrared luminosity function at 60~${\rm \mu m}$.

The LDP model distinguishes the FIRB sources into two categories, \emph{starburst} and \emph{normal} galaxies, and empirically differentiates between their evolutions. AGNs dominate the emitted radiation only at very high luminosities ($L > 2 \cdot 10^{12} L_{\odot}$) and have a negligible influence at the wavelength of interest for OLIMPO.

Given a source at redshift $z$, its flux $S_{\nu}$ at a given wavelength $\lambda=\lambda_{0}$ can be written as a function of the rest frame luminosity $L_{\nu}$ (expressed in W/Hz)
\begin{equation}
	S_{\nu}(L, z, \lambda=\lambda_{0}) = \frac{(1+z)\, K(L,z)\, L_{\nu}(\lambda=\lambda_{0})}{4\pi\, D_{L}^{2}} ,
\end{equation}
where $D_{L}$ is the luminosity distance and $K(L,z)$ is the \emph{K-correction} factor, defined as
\begin{equation}
	K(L,z) = \frac{L_{\nu(1+z)}}{L_{\nu(z=0)}} .
\end{equation}
The number of sources per solid angle and redshift interval is
\begin{equation}
	\frac{dN}{dz\, d\log L}(L,z) = N_{0}(L,z)\, (1+z)^{3}\, \frac{dV}{dz} ,
\end{equation}
where $dV/dz$ is the differential element of volume, once the cosmological parameters are fixed, $N$ is the number of sources per unit volume and luminosity interval as a function of redshift, and $N_{0}$ is given by the luminosity function.

Differential and integrated counts for a given flux $S$ and at wavelength $\lambda=\lambda_{0}$ can be written as
\begin{equation}
	\frac{dN}{dS} = \int_{L} \int_{z} \frac{dN}{dz\, d\log L}(L,z)\, \frac{dz}{dS}(L,z)\, d\log L ,
\end{equation}
\begin{equation}
	N(>S)=\int \frac{dN}{dS}\, dS .
\end{equation}
Hence, the FIRB intensity $I_{\mathit{FIRB}}$ produced by sources with flux $S < S_{\mathit{Max}}$, expressed in Jy/sr, is
\begin{equation}
	I_{\mathit{FIRB}} = \int_{0}^{S_{\mathit{Max}}}S\, \frac{dN}{dS}\, dS ,
\end{equation}
while the fluctuation densities caused by the sources below the observational flux limit $S_{0}$, which correspond to both the confusion noise and sensitivity limit of the experiment, are given by
\begin{equation}
	P_\mathit{fluc}= \int_{0}^{S_{0}} S^{2}\, \frac{dN}{dS}\, dS ,
\end{equation}
thus $P_\mathit{fluc}$ is expressed in Jy$^{2}$/sr. In the case of OLIMPO, the 5-$\sigma$ confusion noise $S_{0}$ ranges from 1.4 to 11.5~mJy in the four wavebands.

\subsection{Point sources}\label{sources}

The contribution of point sources to the total signal is mainly that of radio sources in the two lower frequency bands, and resolved infrared sources (the unresolved ones being already taken into account in the FIRB model) in the higher frequency bands. Their contribution to the total signal is negligible with respect to the CMB at low frequencies and the FIRB at higher frequencies. In particular, the \emph{r.m.s.} of the fluctuations due to radio sources is 4\% of the \emph{r.m.s.} due to the CMB in the OLIMPO beam at 143~GHz and 3\% of the \emph{r.m.s.} due to the FIRB in the OLIMPO beam at 450~GHz. However, it is important to include these sources in the simulation, because their presence may alter the results of the component separation (see Sect.~\ref{map_extraction}) and can strongly affect the estimate of the angular power spectra (see Sect.~\ref{dust_firb_ext}) at high multipoles, with an increase of up to 20\% at $\ell \sim 2000$.

We modeled these two kind of sources with a differential source count $dN(S)/dS$ of the form
\begin{equation}\label{eq:dnds}
	\frac{dN(S)}{dS} =
	\left\{ \begin{array}{ll}
		 A_{N}\, S^{- \gamma_{1}} & \mbox{for } S \geq S_{Brk} , \\
		 A_{N}\, S_{Brk}^{\gamma_{2} - \gamma_{1}}\, S^{- \gamma_{2}} & \mbox{for } S \leq S_{Brk} . \\
	\end{array} \right.
\end{equation}

For the radio sources, we used the WMAP catalogue \citep{2009ApJS..180..283W} to estimate the parameters of Eq.~\ref{eq:dnds} at 143~GHz, obtaining $\gamma_{1} = 0.23$, $\gamma_{2} = 1.03$, $S_{Brk} = 1.9 \times 10^{-3}$~mJy, and the normalization factor $A_{N} = 59$.

For the infrared sources, we fitted the $dN(S)/dS$ from the LDP model with the previous analytical expression \citep{2004MNRAS.348..737F}, deriving $\gamma_{1} = 3.3$, $\gamma_{2} = 1.8$, $S_{Brk} = 1.9$~mJy, and $A_{N} = 18 \cdot 10^{9}$.

\section{Map generation}\label{map_generation}

We generated $ 10^{\circ}\times 10^{\circ} $ maps, by taking into account the instrument characteristics and using as a baseline the OLIMPO experiment (see Sect.~\ref{experiment}). A beam of $5'$ FWHM was used for all the wavebands. This means that we degraded the map angular resolution for the high frequency channels, so that all maps are immediately comparable. We assumed arrays of 19 detectors in the two lower frequency bands (143 and 217~GHz) and 24 detectors in the others. We decided to integrate for 100~h over the observed area of the sky, which is centered on the Galactic coordinates $b = 65^\circ$, $l = 155^\circ$. These assumptions are consistent with a flight of $\sim 15$ days departing from the Svalbard islands (Norway) in June$-$July and circumnavigating the north pole \citep{2008MmSAI..79..792P}.

\begin{figure*}[tbh]
	\centering
	\includegraphics[width=\textwidth, keepaspectratio]{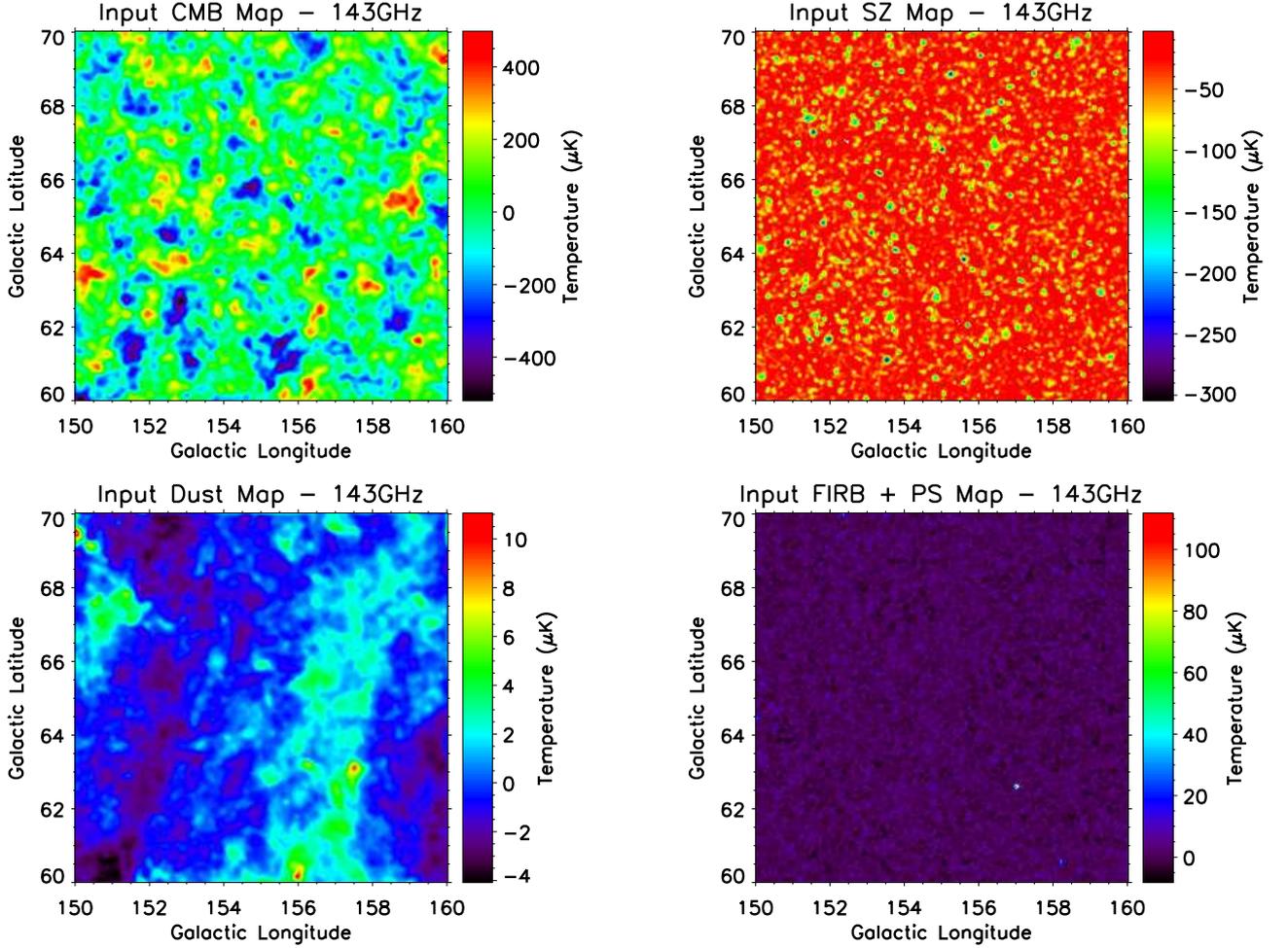}
	\caption{Input maps of the four components at 143~GHz. Signals are expressed in ${\rm \mu K}$. At this frequency, apart from the largest clusters that produce an intense Sunyaev-Zel'dovic effect, the main contribution is due to CMB anisotropies.}\label{fig:input:maps}
\end{figure*}

\subsection{Cosmic microwave background map}

The CMB map is a Gaussian random realization with a null average value and an angular power spectrum that was measured by WMAP and is defined by the cosmological parameters of the $\Lambda$CDM model \citep{2009ApJS..180..306D}. We generated a full sky map using the \emph{synfast} program (part of the \emph{HEALPix} package of \cite{2005ApJ...622..759G}): it was generated for ${\it nside} = 2048$, which roughly corresponds to a pixel dimension of $5/3'$. Hence, each map's pixel is identified by the Galactic coordinates (latitude $b$ and longitude $\ell$) and its flux is $\Delta_{\mathit{CMB}}$. Then, using the \emph{smoothing} program included in the \emph{HEALPix} package, the map is convolved with the instrument beam working in harmonic space. In Galactic coordinates $(b, \ell)$, the convolved map $S_{\it CMB}$ is given by
\begin{equation}\label{eq:cmb_conv}
	S_{\it CMB}(b, \ell) = \int_0^{2\pi} \int_{-\pi/2}^{\pi/2}
		\Delta_{\it CMB}(b' , \ell')\, f(\gamma)\, \cos b'\, db' d\ell' ,
\end{equation}
where $\Delta_{CMB}(b', \ell')$ is the input CMB map towards the Galactic coordinates $(b', \ell')$,
\begin{equation}\label{eq:beam_1}
	f(\gamma) = {1 \over 2\pi \, \sigma_{\it FWHM}^2} \exp \left[ - {1 \over 2}
		\left( {\gamma \over \sigma_{\it FWHM}} \right)^2 \right]
\end{equation}
is a symmetric Gaussian angular response with $\sigma_{\it FWHM} = 5' / \sqrt{8 \ln 2}$ and
\begin{equation}\label{eq:beam_2}
	\gamma = \arccos( \sin b\, \sin b' + \cos b\, \cos b'\, \cos (\ell - \ell') )
\end{equation}
is the angular distance between direction $(b, \ell)$ and direction $(b', \ell')$. Finally, we extracted a $10^\circ \times 10^\circ$ portion of sky. The final map used for the analysis is presented in the top-left panel of Figure~\ref{fig:input:maps}.

\subsection{Sunyaev-Zel'dovich effect map}

We calculated the SZ spectrum in the non-relativistic limit (i.e.\ in the Kompaneets approximation, \citealt{kompaneets}). In this case, the spectrum of all the clusters present in the map is insensitive to the gas temperature: as shown by Eqs.~\ref{eq:kompaneets} and \ref{eq:compton}, the comptonization parameter is indeed frequency-independent and provides only spatial information about the clusters.

Hence, to simulate the SZ effect, we produced a map of the comptonization parameter $y(b, \ell)$ using hydrodynamical simulations. The SZ spectrum $\Delta_{\mathit{SZ}}$ was then computed in the four experiment's wavebands and converted into K
\begin{eqnarray}
	\Delta_{\mathit{SZ},B} & = & \frac{2(k_{B}T_{\mathit{CMB}})^{4}}{h^{3}c^{2}}
		\int_{0}^{\infty} \frac{10^{6}\, T_{\mathit{CMB}}\, E_{B}(\nu)\, g(\nu)}
		{\left . \frac{\partial B(\nu,T)}{\partial T} \right |_{T=T_{\mathit{CMB}}}}\, d\nu \nonumber \\
	& = & 10^{6}\, T_{\mathit{CMB}} \int_{0}^{\infty} E_{B}(\nu) \cdot \nonumber\\
	& & \cdot \left[ \frac{h\nu}{kT_{\mathit{CMB}}} \cdot \coth \left( \frac{h\nu}{2kT_{\mathit{CMB}}} \right) -4 \right] d\nu ,
\end{eqnarray}
where
\begin{equation}
	g(x) = \frac{x^{4}e^{x}}{(e^{x}-1)^{2} \left[ x \cdot \frac{e^{x}+1}{e^{x}-1}-4 \right]} \
	\mbox{with}\ x=\frac{h\nu}{kT_{\mathit{CMB}}} ,
\end{equation}
and $E_{B}(\nu)$ represents the pass-band filter spectrum of the four channels. We then multiplied $\Delta_{\mathit{SZ}}$ by the map of the comptonization parameter $y(b, \ell)$. Finally, the map is convolved with the beam as done for the anisotropy map (see Eq. \ref{eq:cmb_conv})
\begin{equation}\label{eq:sz_conv}
	S_{\mathit{SZ},B}(b,\ell) = \int_{0}^{2\pi} \int_{-\pi/2}^{\pi/2} \Delta_{\mathit{SZ},B}\
		y(b', \ell ')\, f(\gamma)\, \cos b'\, db' d\ell' .
\end{equation}
Figure~\ref{fig:input:maps}, top-right panel, shows the SZ map obtained at 143~GHz.

\subsection{Galactic dust map}

As described in Sect.~\ref{map_generation}, we selected a sky region centered on the Galactic coordinates $b = 65^\circ$, $ l = 155^\circ$: this area is rich in known clusters, observable during the whole flight and the dust contribution is expected to be very low.

Dust emission is estimated using the Schlegel, Finkbeiner \& Davis model $\# 8$ (see Sect.~\ref{sfd_model}), which is able to rescale the IRAS 100~${\rm \mu m}$ map to the desired frequency. It only requires as input the Galactic coordinates and the frequency, giving as output the sky temperature in $\mu$K: $\Delta_{\it Dust}(b,\ell,\nu)$. It is then convolved with the filter transfer function and the instrument beam $f(\gamma)$ (see Eqs.~\ref{eq:beam_1} and \ref {eq:beam_2})
\begin{eqnarray}
	S_{\it Dust,B}(b,\ell) & = & \int_{0}^{2 \pi} \int_{-\pi/2}^{\pi/2}
		f(\gamma)\, \cos b'\, db' d\ell' \cdot \nonumber \\
		& & \cdot \int_{0}^{\infty} \Delta_{\it Dust}(b',\ell',\nu)\, E_{B}(\nu)\, d\nu .
\end{eqnarray}
Figure~\ref{fig:input:maps}, bottom-left panel, presents the Galactic dust map at 143~GHz.

\begin{figure*}[htb]
	\centering
	\includegraphics[width=\textwidth, keepaspectratio]{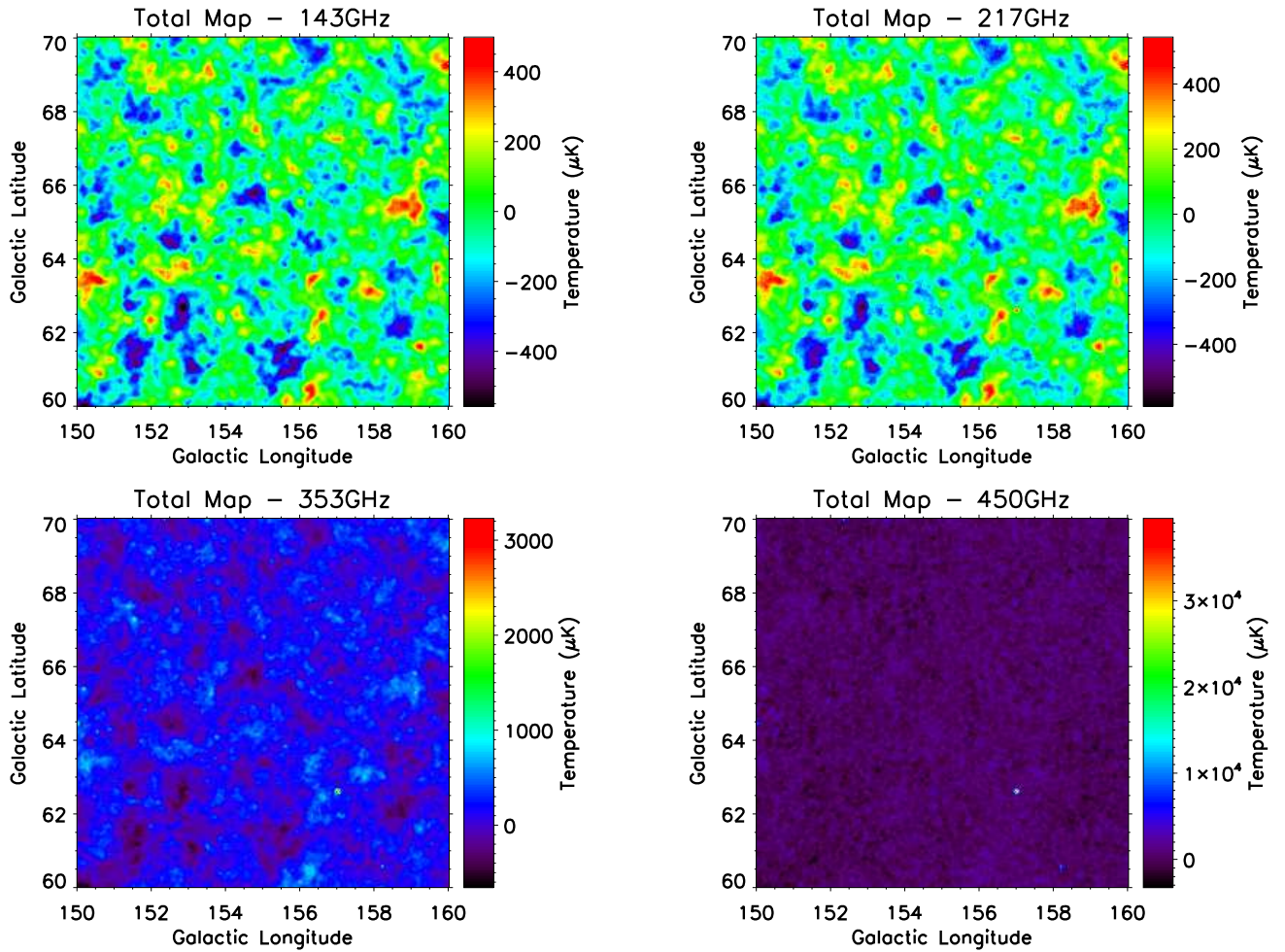}
	\caption{Total maps predicted at the four frequency bands of OLIMPO: these are the sum of the CMB anisotropies, SZ effect, dust, and FIRB components plus the radio and infrared sources and the expected noise in the four bands. Signals are expressed in ${\rm \mu K}$. We note that the CMB anisotropy component dominates the 143 and 217~GHz channels, while the 353 and 450~GHz channels are dominated by Galactic dust. The far infrared background contribution is comparable to the Galactic dust one, but seeing it is more difficult since it lacks of structures at scales larger than beam size. However, when adding the point sources, these loom over the other components, particularly at 450~GHz.}\label{fig:total:maps}
\end{figure*}

\subsection{FIRB map}

Our FIRB maps were compiled in two stages. In the first step, we produced four maps in the four OLIMPO wavebands using a Gaussian distribution, $\mathcal{N}(\Delta_{\mathit{FIRB}}, \sigma_{\mathit{FIRB}}, b, \ell)$, with the average and the standard deviation provided by the LDP model in Jy/sr and ${\rm Jy/ \sqrt{sr}}$, respectively. The total signal was then converted into ${\rm \mu K}$.

In the second step, we randomly distributed the radio and infrared point sources over the maps, following the differential source counts $dN(S)/dS$ described in Sect.~\ref{sources}. Each radio source was scaled to higher frequency using a power law $f(\nu) \propto \nu^{-\alpha}$, where the spectral index $\alpha$ varies between -2.5 and 0.5 \citep{2001ApJ...562...88S}. For the infrared sources, we distributed only sources with a flux 3-$\sigma_{\mathit{FIRB}}$ above the FIRB background, to consider only sources that are not confused (hence already taken into account as part of the background).

Finally, each map is convolved with the instrument beam: the resulting map at 143~GHz is shown in Figure~\ref{fig:input:maps}, bottom-right panel. We note that we did not include any clustering component in our simulations: this certainly affects the input FIRB signal \citep{2005ApJ...621....1G}, but not the results discussed in Sect.~\ref{dust_firb_ext}. Adding the clustering would increase the FIRB power spectrum at high multipoles, but the ability to disentangle the different components remains the same (and so the ability in recovering the input spectrum), as long as a pixel-based map extraction method is used (see Sect.~\ref{map_extraction}).

\subsection{Noise map}

In building our model, we made no attempt to simulate systematic effects, and we took into account only white noise that has a null average and standard deviation equal to
\begin{equation}\label{eq:delta_noise}
	\Delta_{Noise,B} = \frac{\mathit{NET}_{B}\, N_{Pixel}}{\sqrt{T_{Int}N_{Bolo,B}}} ,
\end{equation}
where $\mathit{NET}_{B}$ is the bolometer's noise in ${\rm \mu K /\sqrt{Hz}}$ and ranges from $\sim 150\, {\rm \mu K / \sqrt{Hz}}$ at 143~GHz to $\sim 550\, {\rm \mu K /\sqrt{Hz}}$ at 450~GHz, $N_{Bolo,B}$ is the number of receivers in each channel, $T_{Int}$ is the integration time, and $N_{Pixel}$ is the number of pixel in a row (or column) of the map. Each map's pixel has a dimension equal to \emph{FWHM}/3, thus
\begin{equation}
	\frac{N_{Pixel}}{\sqrt{T_{Int}}} = \frac{3 \cdot 60' \cdot \, {\rm deg}}{5' \cdot \sqrt{\rm 3600\, s \cdot \, {\rm deg^{2}}}}
	= \frac{3}{5}\frac{1}{\sqrt{\rm s}} ,
\end{equation}
where we have taken into account that we are integrating for 1~h over a 1~${\rm deg^2}$ region. As a consequence, Eq.~\ref{eq:delta_noise} could be rewritten as
\begin{equation}
	\Delta_{Noise,B} = \frac{3}{5} \frac{\mathit{NET}_{B}}{\sqrt{N_{Bolo,B}}} .
\end{equation}
The computed noise is finally convolved with the instrument beam as done in Eqs.~\ref{eq:cmb_conv} and \ref{eq:sz_conv}.

\begin{figure*}[htb]
	\centering
	\includegraphics[width=\textwidth, keepaspectratio]{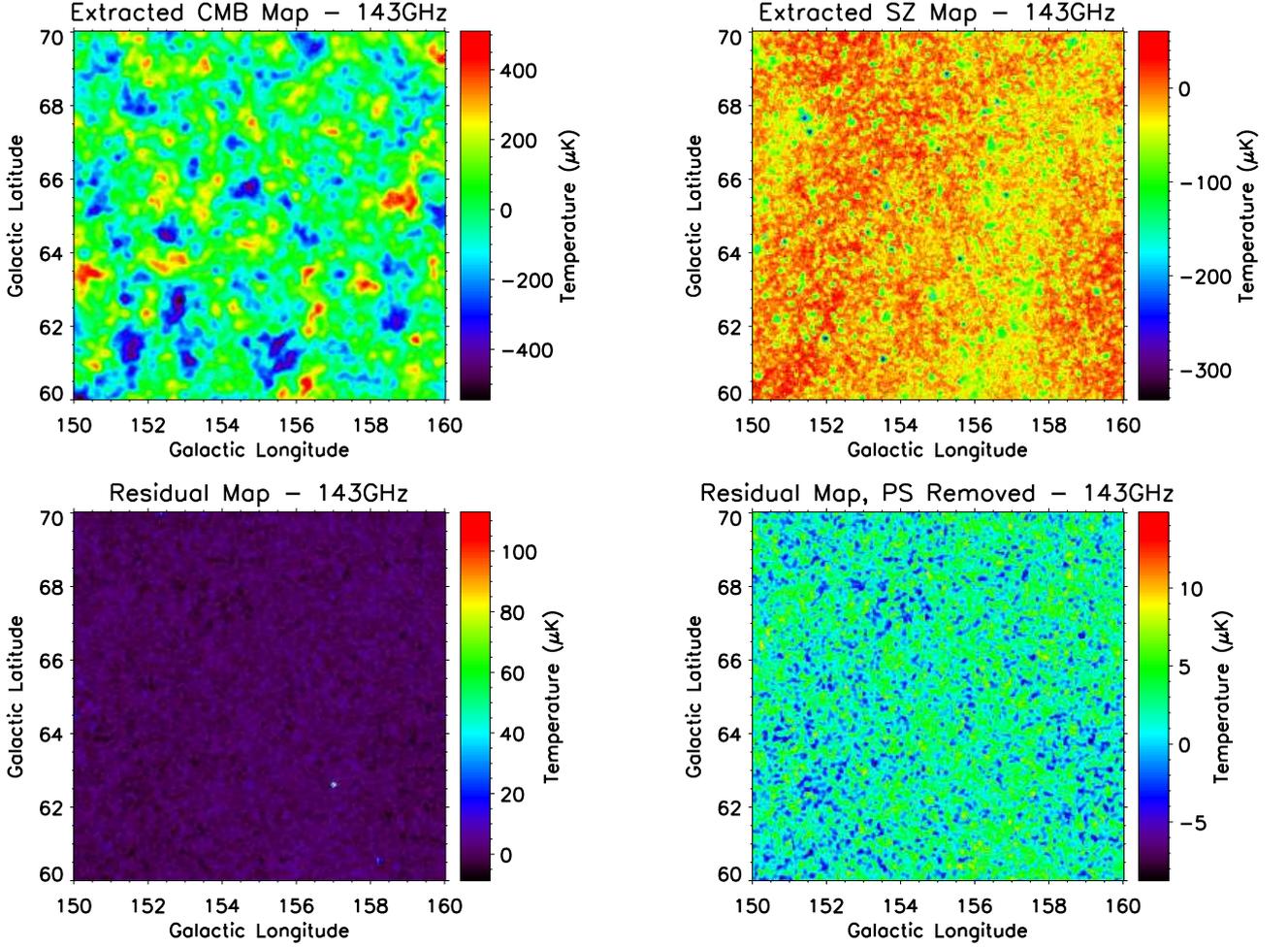}
	\caption{Extracted maps of the four components at 143~GHz. Signals are expressed in ${\rm \mu K}$. The anisotropy map is the most robustly extracted; the SZ map clearly shows the brighter clusters; the dust and FIRB component cannot be easily distinguished because their spectra are too similar.}\label{fig:ext:maps}
\end{figure*}

\section{Map extraction}\label{map_extraction}

The total map for each frequency band is obtained by summing the maps of each component (see Figure~\ref{fig:input:maps}): the results are the four maps shown in Figure~\ref{fig:total:maps}.

The signal caused by Sunyaev-Zel'dovich effect is always subdominant: the cosmic microwave background anisotropies dominate the two lower channels, while at 353 and 450~GHz the far infrared background and the Galactic dust contributions are of comparable strength. FIRB is not evident in the maps because it lacks structure on scales larger than the beam size. However, when adding individual point sources, these stand out with respect to of the other components, particularly at 450~GHz. A comprehensive comparison of the different component-separation methods is reported in \cite{2008A&A...491..597L}. However, we decided to use a simple $\chi^2$ approach to ensure a tighter control on each step of our analysis. This method only requires an \emph{a priori} knowledge of the component frequency spectrum but not of their spatial distribution.

The total signal can be expressed in terms of the input signal and its frequency spectrum using the matrix formalism
\begin{equation}\label{eq:matrix_add}
	\mathbf{S_{T}} (b,\ell) = \mathbf{M} \cdot \mathbf{S_{C}} (b,\ell) + \mathbf{S_{N}} (b,\ell) ,
\end{equation}
where $\mathbf{S_{T}}$ is a vector containing the total maps in the four frequency bands, $\mathbf{M}$ is a $4 \times 4$ matrix whose columns comprise the scaling factors for each of the four components relative to one frequency band of choice, $\mathbf{S_{C}}$ is a vector consisting of the four component maps computed at the chosen frequency, and $ \mathbf{S_{N}}$ contains the four noise maps.

To disentangle the four components and obtain their maps at the chosen frequency, we used the $\chi^2$ method. The optimal estimation of the input signals was obtained by minimizing, for each pixel, the quantity \citep{2002MNRAS.330..497D}
\begin{equation}\label{eq:matrix_chi}
	\chi^2 = (\mathbf{S_{T}} - \mathbf{M} \cdot \mathbf{S_{C}})^\dag \cdot \mathbf{N}^{-1}
		(\mathbf{S_{T}} - \mathbf{M} \cdot \mathbf{S_{C}}) ,
\end{equation}
where $\mathbf{N}$ is the noise covariance matrix, which in our case is reduced to a diagonal matrix whose elements are the noise variances at the four frequencies. The solution to Eq.~\ref{eq:matrix_chi} is
\begin{equation}
	\mathbf{\tilde{S}_{C}} (b,\ell) = \left [ \mathbf{M}^T \mathbf{N}^{-1} \mathbf{M} \right ]^{-1}
		\mathbf{M}^T \mathbf{N}^{-1}\, \mathbf{S_{T}} (b,\ell) ,
\end{equation}
where $\mathbf{\tilde{S}_{C}}$ is our optimal estimation of $\mathbf{S_{C}}$, i.e.\ it contains the signal of the component extracted from the four total maps.

\section{Results}\label{results}

While the CMB and SZ frequency spectra are theoretically and experimentally well constrained, this is not true for the other components. As a consequence, we decided to extract the CMB and SZ signals and leave the dust, FIRB, and sources in a \emph{residual} map. The results of the component separation are shown in Figure~\ref{fig:ext:maps}, where we chose to extract the components at 143~GHz.

\subsection{CMB}

We computed the angular power spectrum on the extracted CMB map using \emph{anafast}, part of the \emph{HEALPix} package \citep{2005ApJ...622..759G}. However, appropriate corrections must be made because we observed only a small fraction of the sky \citep{2002ApJ...567....2H}: these include those for instrumental beam shape, scanning strategy, and instrumental noise. Since some of these characteristics will be known only once the experiment have flown, we used a simpler approach. We computed the power spectra of both the input map and a full sky CMB one, using the same parameters to generate them: the ratio of the two power spectra then provided us with the correction factor for all the spectra computed in the partial maps.

\begin{figure}[htb]
	\centering
	\includegraphics[width=8cm, keepaspectratio]{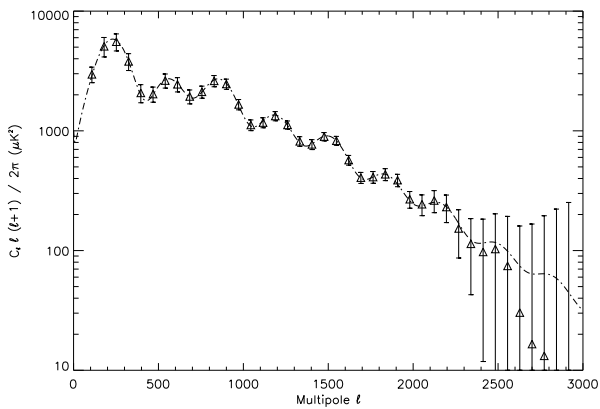}
	\caption{CMB angular power spectrum used to generate the input map (dot-dashed line) compared to the recovered spectrum (triangles): we used a binning of $\Delta \ell = 4\pi / b_{\it Max} = 72$, $b_{\it Max}$ being the map size.}\label{cmb_spectra}
\end{figure}

Figure~\ref{cmb_spectra} shows the angular power spectrum used to generate the input CMB map (dot-dashed line) compared to the spectrum computed on the extracted CMB map (triangles), using a binning of $\Delta \ell = 4\pi / b_{\it Max} = 72$, where $b_{\it Max}$ represents the map size. Thanks to OLIMPO's characteristics (high resolution, low noise receivers, and wide frequency range), we are able to fully recover the spectrum up to $\ell \simeq 2200$; at higher multipoles, our data still fit the input spectrum, but the beam contribution is more prominent increasing the error in the $C_\ell$ estimation.

\subsection{SZ effect}

\begin{figure}[htp]
	\centering
	\includegraphics[width=8cm, keepaspectratio]{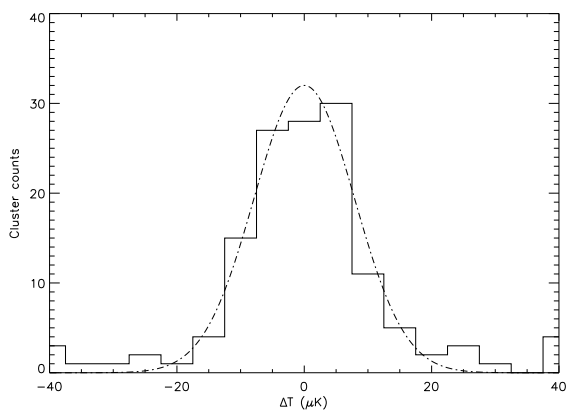} 
	\caption{Flux difference of the cluster found in the input and in the extracted SZ maps. Data dispersion is well fitted by a Gaussian distribution centered on zero and with standard deviation $\sigma = 7.9\, {\rm \mu K}$.}\label{diff_flussi}
\end{figure}

We identified galaxy clusters in the input and extracted SZ maps, using the \emph{SExtractor} algorithm \citep{1996A&AS..117..393B}, choosing a threshold $\sigma_{\it Thr} = 3\, \sigma_{\it SZ}$, where $\sigma_{\it SZ}$ is the estimated standard deviation of the SZ map's background. We found $ \sim 270$ clusters in the input map and $\sim 170$ in the extracted one, without any false detection (i.e.\ clusters found in the extracted map but not present in the original one). We then computed the difference between the fluxes of the clusters found in both maps: these data have a distribution similar to that of a Gaussian centered on zero and with a standard deviation $\sigma = 7.9\, {\rm \mu K}$ (see Figure~\ref{diff_flussi}).

\begin{figure}[htp]
	\centering
	\includegraphics[width=8cm, keepaspectratio]{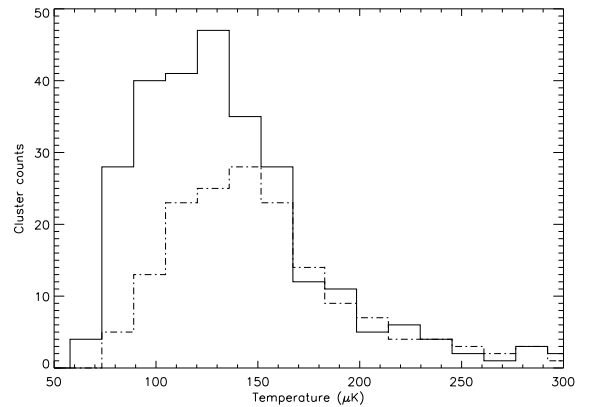}
	\caption{Comparison of the number of clusters found per binned flux in the input (solid line) and extracted (dot-dashed line) SZ maps. The binning is twice the standard deviation of the flux difference distribution, i.e.\ $\Delta T = 15.8\, {\rm \mu K}$.}\label{conteggi_cluster}
\end{figure}

We also compared the number of clusters found per binned flux in both maps, reporting our results in Figure~\ref{conteggi_cluster}: the binning is twice the previous standard deviation, i.e.\ $\Delta T = 15.8\, {\rm \mu K}$. We recovered all the clusters brighter than $\sim160\, {\rm \mu K}$, while we lost about $ 50\%$ of the fainter ones due to the lower signal-to-noise ratio. 

\begin{figure*}[htb]
	\centering
	\includegraphics[width=8cm, keepaspectratio]{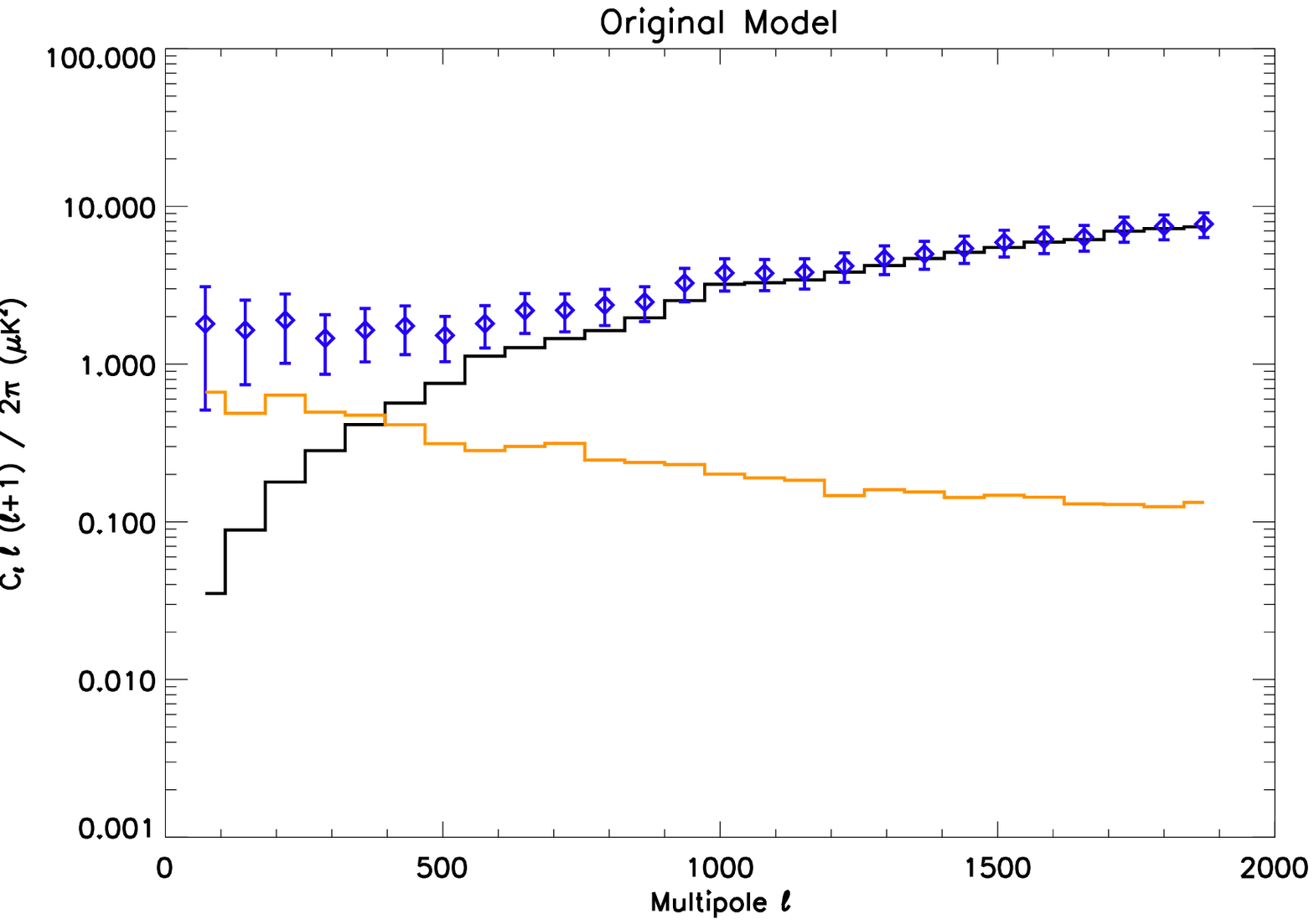}
	\includegraphics[width=8cm, keepaspectratio]{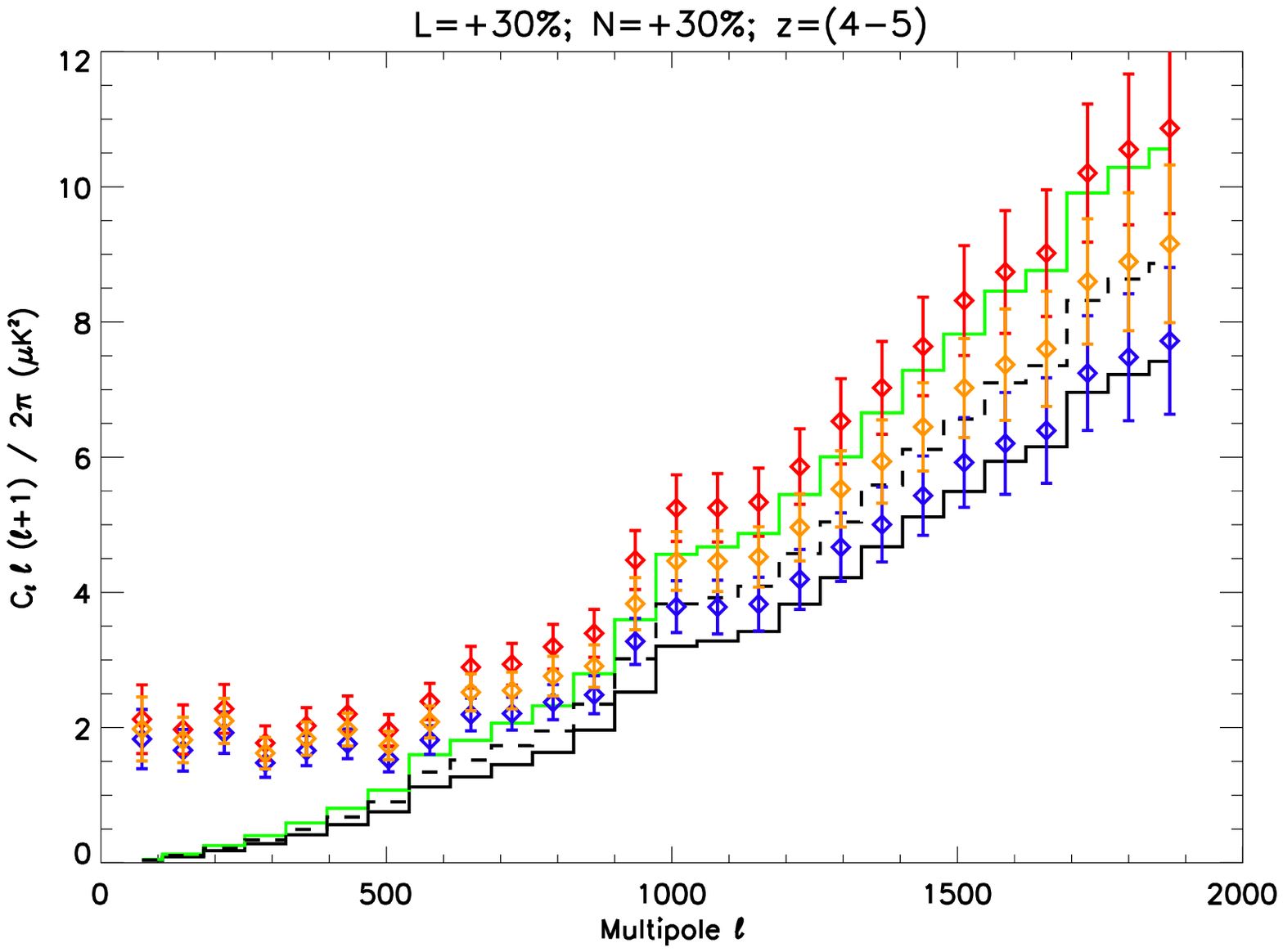}
	\caption{Angular power spectra obtained on the input maps of dust and FIRB and on the extracted maps. On the left, the yellow line is the dust spectrum computed on the input dust-only map; the black line represents the FIRB spectrum computed on the input map, using the original LDP model. The blue diamonds are the binned spectra computed on the extracted map. On the right, the comparison between the original LDP model and the modified ones when incrementing by 30\% the luminosity or the number of sources in the redshift range $z = 4 - 5$. The dashed line illustrates the effect of increasing the luminosity of sources; the green line corresponds to an augmentation of 30\% in the number of sources in the same redshift range; the solid line is the original FIRB spectrum. The yellow and red diamonds are the extracted spectrum for the two modified models, while the blue diamonds are for the original one. The error bars represent the 1-$\sigma$ uncertainty in the computed binned spectra.}\label{final_spectra}
\end{figure*}

\subsection{Dust \& FIRB}\label{dust_firb_ext}

The extracted map contains the contribution of Galactic dust, FIRB, and point sources. Point sources must be removed otherwise the angular power spectra can become distorted at high multipoles. We achieved this using \emph{SExtractor} \citep{1996A&AS..117..393B} with a threshold $\sigma_{\it Thr} = 3\, \sigma_{\it FIRB}$, where $\sigma_{\it FIRB}$ is the standard deviation of the extracted map's background as estimated by the program itself. We then computed the angular power spectrum of the input dust and FIRB maps, separately, and of the extracted map: the left plot of Figure~\ref{final_spectra} shows the dust spectrum (yellow line), the FIRB spectrum (black line), and the spectrum obtained for the extracted map (blue diamonds). The plot shows that dust is dominant on large scales ($\ell \lesssim 200$), but becomes negligible at higher multipoles.

\begin{figure}[htb]
	\centering
	\includegraphics[width=8cm, keepaspectratio]{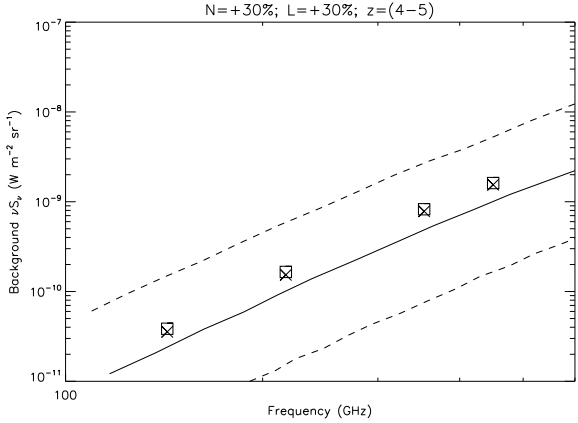}
	\caption{Comparison between the analytic form of the infrared background measured by FIRAS \citep{1998ApJ...508..123F} and the modified FIRB models produced by incrementing by 30\% the luminosity (crosses) or the number of sources (boxes) in the redshift range $z = 4 - 5$. The solid and dashed lines represent the most accurate estimation and the error at 1-$\sigma$ in the FIRB spectrum obtained with FIRAS; the four boxes and crosses are the fluxes of the modified FIRB model at 143, 217, 353, and 450~GHz.}\label{firas}
\end{figure}

We then performed the same analysis after changing the FIRB model. In this way, we tested whether OLIMPO is able to discriminate between the different evolutionary models:
\begin{itemize}
	\item we considered one redshift interval $\Delta z =1$ in the range $z=0 - 7$ and increased the number of sources with respect to the original LDP model: this simulated a higher star formation rate in this redshift interval;
	\item we considered one redshift interval $\Delta z = 1$ in the range $z=0 - 7$ and increased the luminosity of sources with respect to the original analysis, while the number of sources remained the same.
\end{itemize}
In the first case, we modified the number of sources, their luminosities, and their fluxes: increasing the number of sources in a given redshift interval indeed caused the total luminosity and flux contribution from the same interval also to be augmented. 

We tried different increments, focusing our tests in the range $z = 4 - 5$, where the star formation rate is not well known. For each increment in luminosity or number of sources, the corresponding modified FIRB signal was computed again in each frequency band, added to the other components, and extracted together with the Galactic dust. We computed once more the angular power spectra of both the modified FIRB input map and the extracted one. We then estimated our ability to disentangle the different models calculating the Bayes factor $K_{{\it Orig} / {\it Mod}} = p(C_\ell | M_{\it Orig}) / p(C_\ell | M_{\it Mod})$, where $p(C_\ell | M_x)$ is the marginal likelihood for model $M_x$ (i.e.\ the original one $M_{\it Orig}$ or a modified one $M_{\it Mod}$) and is given by
\begin{equation}\label{eq:bayes}
	p(C_\ell | M_x) \propto \exp \left[ -\frac{1}{2} \cdot \sum_\ell \left(
		\frac{C_\ell^{\it x} - C_\ell^{\it Exp}}{\sigma_{C_\ell}} \right)^2 \right] .
\end{equation}
The proportional sign is indicated because of the absence of the normalization factor, which vanishes when computing the ratio, $C_\ell^{\it Exp}$ is the binned angular power spectrum obtained on the extracted map, while $C_\ell^{\it x}$ is the one computed on the input FIRB-only map.

The smallest increment for which we were able to disentangle the original and modified FIRB models corresponds to an increase of 30\% in the luminosity and of 20\% in the number of sources. When comparing $C_\ell^{\it Exp}$ (using as input the original model) to $C_\ell^{\it Orig}$ and $C_\ell^{\it Mod}$, we obtained $K_{{\it Orig} / 1.3 L} = 35$ when augmenting the luminosity and $K_{{\it Orig} / 1.2 N} = 87$ when incrementing the number of sources. Since the Bayes factors is higher than 10, there is strong support for the original LDP model, which was indeed the input spectrum of $C_\ell^{\it Exp}$.

The right plot in Figure~\ref{final_spectra} shows the results obtained after increasing by 30\% both the luminosities (dashed line and orange diamonds represent original and extracted spectra respectively) and the number of sources (green line and red diamonds, original and extracted spectra respectively) in the range $z = 4 - 5$, compared to the original model (solid line and blue diamonds, original and extracted spectra, respectively). In all cases, we were able to recover the correct input spectrum.

We note that the modified models are consistent with the limits imposed by FIRAS data \citep{1998ApJ...508..123F} in all the redshift intervals at a 1-$\sigma$ level: Figure~\ref{firas} reports the case of increasing by 30\% both the luminosity (crosses) and the number of sources (squares) in the range $z = 4 - 5$.

\section{Conclusions}\label{conclusions}

We have presented accurate simulations of the expected sky brightness in the OLIMPO wavebands. Using a simple component-separation method, we have efficiently disentangled the input signals.

The CMB angular power spectrum is efficiently recovered up to $\ell \simeq 3000$: even if OLIMPO is not specifically designed to study the CMB anisotropies, this gives us an idea of its capabilities. Moreover, the achieved result allows us to place stronger constrains on the damping tail of the CMB spectrum.

Using the Sunyaev-Zel'dovich effect, we have been able to extract most of the galaxy clusters present in our input map: we are complete down to 160 ${\rm \mu K}$ and detect 50\% of the fainter clusters, giving a 60\% efficiency overall.

The separation of the dust and FIRB components was inefficient at the pixel level. However, after removing the radio and infrared point sources, the computation of the angular power spectra on the extracted maps was found to be in agreement with the sum of the two input components. Moreover, the dust's spectrum quickly declines dramatically and its contribution becomes negligible on scales smaller than $\sim 1^\circ$ ($\ell \gtrsim 200$). Even if we had not included any clustering component, which should increase the input FIRB spectrum at high multipoles, this would not affect our ability to disentangle the different components and recover their input spectra.

We were also able to place stronger constrains on far infrared background models: we were able to distinguish between the original LDP model and a modified one, in which we had increased the number of sources by 20\% or their luminosities by 30\% in a given redshift interval $\Delta z = 1$.

Hence, thanks to its characteristics (high angular resolution, several detectors each having low noise, long duration flight covering a selected patch of the sky with a long integration time), OLIMPO will be able to achieve its scientific goals, complementing on the one hand the shallower survey of Planck, and on the other the lower frequency surveys of bolometric arrays on large-size ground-based telescopes.

\acknowledgements

The authors would like to thank all the people of the Experimental Cosmology Group at the Physics Dept., University of Rome ``La Sapienza''. In particular, we would like to mention Drs.~F.~Piacentini and E.~Valiante for their useful suggestions. We are very grateful to Dr.~S.~Sadeh from Tel Aviv University for providing us with the comptonization parameter map obtained from accurate hydrodynamical simulations. This work has been supported by Italian Space Agency contracts ``COFIS'' and ``OLIMPO'' and by PRIN 2006 ``Cosmologia Millimetrica con Grandi Mosaici di Rivelatori'' of the Ministero dell'Istruzione, dell'Universit\`{a} e della Ricerca. We also acknowledge support from the Faculty of the European Space Astronomy Centre (ESAC). The results presented in this paper have been derived using the \emph{HEALPix} package \citep{2005ApJ...622..759G} and WMAP data \citep{2009ApJS..180..306D}.

\bibliographystyle{aa}
\bibliography{12838}

\end{document}